\documentclass[useAMS,usenatbib]{mn2e}

\usepackage{amsmath}
\usepackage{amssymb}
\usepackage{soul}
\usepackage{natbib}
\usepackage{graphicx}
\usepackage{caption}
\usepackage{subcaption}
\usepackage{enumerate}
\usepackage{datetime}
\usepackage{color}
\usepackage{hyperref}
\usepackage{draftcopy}

\definecolor{dylan}{rgb}{0.431,0.106,0.537}

\begin{document}

\title[Ablation of continuum optically thick discs]
{Line-driven ablation of circumstellar discs: III. Accounting for and analyzing the effects of continuum optical depth}

\author[N. D. Kee et al.]
{Nathaniel Dylan Kee$^1$\thanks{Email: nathaniel-dylan.kee@uni-tuebingen.de},
Stanley Owocki$^2$, and Rolf Kuiper$^1$\\
 $^1$ Institut f\"ur Astronomie und Astrophysik, Eberhard Karls Universit\"at T\"ubingen, D-72076 T\"ubingen, Germany\\
 $^2$ Bartol Research Institute, Department of Physics and Astronomy,
 University of Delaware, Newark, DE 19716, USA\\ 
 }

\def\<<{{\ll}}
\def\>>{{\gg}}
\def\wig{{\sim}}
\def\spose#1{\hbox to 0pt{#1\hss}}
\def\+/-{{\pm}}
\def\=={{\equiv}}
\def\mubar{{\bar \mu}}
\def\mustar{\mu_{\ast}}
\def\Lambar{{\bar \Lambda}}
\def\Rstar{R_{\ast}}
\def\Mstar{M_{\ast}}
\def\Lstar{L_{\ast}}
\def\Tstar{T_{\ast}}
\def\gstar{g_{\ast}}
\def\vth{v_{th}}
\def\grad{g_{rad}}
\def\glines{g_\mathrm{lines}}
\def\Mdot{\dot M}
\def\mdot{\dot m}
\def\yr{{\rm yr}}
\def\ksec{{\rm ksec}}
\def\kms{{\rm km/s}}
\def\qad{\dot q_{ad}}
\def\qlines{\dot q_\mathrm{lines}}
\def\solar{\odot}
\def\Msun{M_{\solar}}
\def\msbyr{\Msun/\yr}
\def\Rsun{R_{\solar}}
\def\Lsun{L_{\solar}}
\def\Be{{\rm Be}}
\def\Rpole{R_{\ast,p}}
\def\Req{R_{\ast,eq}}
\def\Rmin{R_{\rm min}}
\def\Rmax{R_{\rm max}}
\def\Rstag{R_{\rm stag}}
\def\vinf{V_\infty}
\def\Vrot{V_{rot}}
\def\Vcrit{V_{crit}}
\def\d{\mathrm{d}}
\def\rres{\mathbf{r}_\mathrm{res}}
\def\ro{\mathbf{r}_\mathrm{loc}}
\def\rs{\mathbf{r}_\mathrm{s}}
\def\req{\mathbf{r}_\mathrm{eq}}

\maketitle

\begin{abstract}
In studying the formation of massive stars, it is essential to consider the strong radiative feedback on the stars' natal environments from their high luminosities ($10^4 \sim 10^6 L_\odot$).
Given that massive stars contract to main-sequence-like radii before accretion finishes, one form this feedback takes is UV line-acceleration, resulting in outflows much like those expected from main-sequence massive stars.
As shown by the prior papers in this series, in addition to driving stellar winds, 
such line forces also ablate the surface layers off of circumstellar discs within a few stellar radii of the stellar photosphere.
This removal of material from an accretion disc in turn results in a decreased accretion rate onto the forming star.
Quantifying this, however, requires accounting for the continuum optical depth of the disc along the non-radial rays required for the three-dimensional line-acceleration prescription used in this paper series.
We introduce the ``thin disc approximation'', allowing these continuum optical depths arising from an optically thick but geometrically thin disc to be dynamically treated in the context of radiation-hydrodynamics simulations.
Using this approximation in full dynamical simulations, we show that such continuum optical depth effects only reduce the disc ablation by 30 percent or less relative to previous simulations that ignored continuum absorption.
\end{abstract}

 
\section{Introduction}
\label{sec:intro}

Perhaps the most archetypical scenario in which discs arise around stars is during star formation.
These accretion discs have long been observed around solar and later type stars, but confirming their presence around forming luminous massive stars is observationally much more challenging 
\citep[see, e.g.][]{ZinYor07, CesGal07}.
Additionally, the immense feedback imposed on the circumstellar environment by the forming massive star means that its accretion disc will tend to be disrupted relatively quickly after the star reaches the main sequence.
Nevertheless, numerical modeling of massive star formation has shown that accretion through such a disc can form quite massive stars, either alone \citep{KuiKla10b, KuiKla11, KuiYor13a, KlaPud16, HarDou17, KuiHos18}, or enhanced by accretion away from the disc plane mediated by so-called ``radiative Rayleigh-Taylor instabilities'' \citep{KruKle09, CunKle11, RosKru16}.
As observational resolving power improves, discs begin to be identified around forming massive stars \citep[see][for an early case]{KraHof10}.
The most recent observations even manage to resolve Keplerian rotation in these discs
\citep[e.g.][]{JohRob15,IleCyg16,BeuWal17,GinBal18,SanKoe18} with several large observing campaigns continuing the search \citep[e.g.][]{CesSan17, BelCes18, BeuMot18}.

Prior theoretical studies of massive star formation have tended to focus on au to pc length scales, under the assumption that disc material that has managed to proceed all the way down to within a few au of the stellar surface without being deflected by stellar feedback is fated to end up in the star.
Additionally, these studies often ignore the effects of UV line-driven stellar winds, arguing that the mass loss in such outflows  is at most a small fraction of the $\sim 10^{-4} - 10^{-3}$ M$_\odot$/yr accretion rates generally invoked for massive star formation.
However, by neglecting the final accretion through the last tens of stellar radii and onto the forming star, these studies neglect the region where UV-line acceleration is the strongest.
It is in this region that disc ablation can be initiated.
This omission of the final accretion, coupled with the results presented in \cite{KeeOwo16a} (hereafter paper I) and \cite{KeeOwo18} (hereafter paper II), motivate us to investigate the role of line-driven disc ablation in sculpting these final miles of the accretion process.

In undertaking such a study, it is no longer feasible to ignore the effects of continuum opacity as we have done when studying classical Be stars.
At the same time, calculating optical depth along the large number of unique, non-radial rays used in calculating the necessary three-dimensional line acceleration is computationally untenable.
As an alternative to such a computationally expensive undertaking, in section \ref{sec:thin_disc} we introduce an approximate, analytic method for calculating optical depths through an optically thick, but geometrically thin disc and test the accuracy of this method by comparing it to long-characteristic ray tracing.
In section \ref{sec:results} we follow this up by presenting results of radiation-hydrodynamics simulations of the line-driven ablation of material from optically thick discs, including now continuum opacity to constrain its role in reducing the rate of ablation of disc mass.
Finally, in section \ref{sec:summary} we sum up the results and conclusions of this paper, as well as presenting potential directions for the research in this series and other future studies.
 
\section{The Thin Disc Approximation}
\label{sec:thin_disc}

\subsection{Derivation of the Thin Disc Approximation}

For the temperatures and densities within a few stellar radii of a forming massive star, continuum electron scattering is a dominant source of opacity.
However, correctly treating scattering radiative transfer requires knowledge of a non-local scattering source function \citep[see, e.g.][]{MihMih84}.
While we were able to approximate this source function for the predominantly wind-located multiple \emph{line} resonances in paper II, here we are concerned with the \emph{continuum} scattering opacity of the circumstellar disc, and such a method is no longer sufficient.
Instead using a Monte Carlo method, it is possible to bypass this problem, but doing so is computationally prohibitive\footnote{Work by \cite{Har15} suggests that this computational cost can be overcome by massively parallel codes, but this method and the computational resources to implement it are not yet widespread.}.
We therefore choose to 
initially consider a pure absorption model (i.e. omitting the source function).
Such a method will systematically over estimate the reduction in continuum flux, and thereby underestimate line-acceleration and the resulting ablation rate.
Omitting the opacity altogether, however, systematically overestimates line-acceleration and the impact of ablation.
By comparing these two limits, we therefore will be able to bracket the true line-acceleration.

In order to implement this pure continuum opacity, we begin with the standard integral equation for optical depth between the position $s_\mathrm{s}$ on the stellar surface and $s_\mathrm{loc}$ at the local point where we wish to calculate line-acceleration

\begin{equation}\label{eqn:tau_s}
\tau \equiv \int_{s_\mathrm{s}}^{s_\mathrm{loc}} \kappa(s) \rho(s) ds\;, 
\end{equation}
where the position dependent opacity, $\kappa(s)$ can be immediately replaced by the constant continuum electron scattering opacity, $\kappa_e$, and the position dependent density, $\rho(s)$ by the function of cylindrical coordinates $\{R,z,\phi\}$,

\begin{equation}\label{eqn:rho_d}
\rho_\mathrm{d}(R,z,\phi) = \rho_\mathrm{eq}(R,\phi) e^{-z^2/(2H(R)^2)}\;,
\end{equation}
where 
$H(R)$ is the disc scale height, 

\begin{equation}\label{eqn:scale_height}
H(R) \equiv \frac{c_s}{V_\mathrm{orb,o}}\frac{R^{3/2}}{R_\ast^{1/2}}\;,
\end{equation}
expressed in terms of stellar radius, $R_\ast$, sound speed, $c_s$, and orbital speed at the stellar surface, $V_\mathrm{orb,o}\equiv \sqrt{G M_\ast/R_\ast}$, further a function of Newton's gravitation constant, $G$, and the stellar mass\footnote{We once again ignore self-gravity of the disc, as this is orders of magnitude weaker than the stellar gravity in the near-star region.}, $M_\ast$.
Note that, while we here allow the equatorial density, $\rho_\mathrm{eq}$, to be an arbitrary function of $R$ and $\phi$ to reflect the evolving hydrodynamical structure through which optical depth is being calculated, we assume that the disc remains in vertical hydrostatic equilibrium.
Comparison with the disc evolution in paper I shows that the un-ablated parts of the circumstellar disc match these criteria well.

Even with a constant opacity and known vertical density variation of the disc, however, in general equation \ref{eqn:tau_s} is not analytic, and will need to be numerically obtained from the hydrodynamic structure.
As mentioned before, non-radial ray tracing for each of the unique rays\footnote{For the simulations included in this paper, the grid has 256 cells in the radial direction and 257 in the latitudinal direction. Each of these cells has 36 unique rays for the line acceleration calculation, for a total of nearly 2.4 million unique rays needing optical depth calculated at each time step.} involved in the line-acceleration prescription is computationally prohibitive.
Thus, we require an approximate method for treating optical depth between the stellar surface and the location where we calculate line-acceleration.
Given that the wind is quite optically thin in the continuum, a standard and well justified approximation in the computation of line-driven stellar winds, this approximate method only needs to account for the optical depth of the disc.
Further, given that the bulk of the disc remains close to vertical hydrostatic equilibrium despite the effects of ablation (see paper I), the vertical stratification of the disc suggests that most of this opacity will come from a relatively narrow region around the disc midplane.
Based on these assumptions, we propose what we refer to for the remainder of this paper as the ``thin disc approximation".

\begin{figure}
\includegraphics[width=0.5\textwidth]{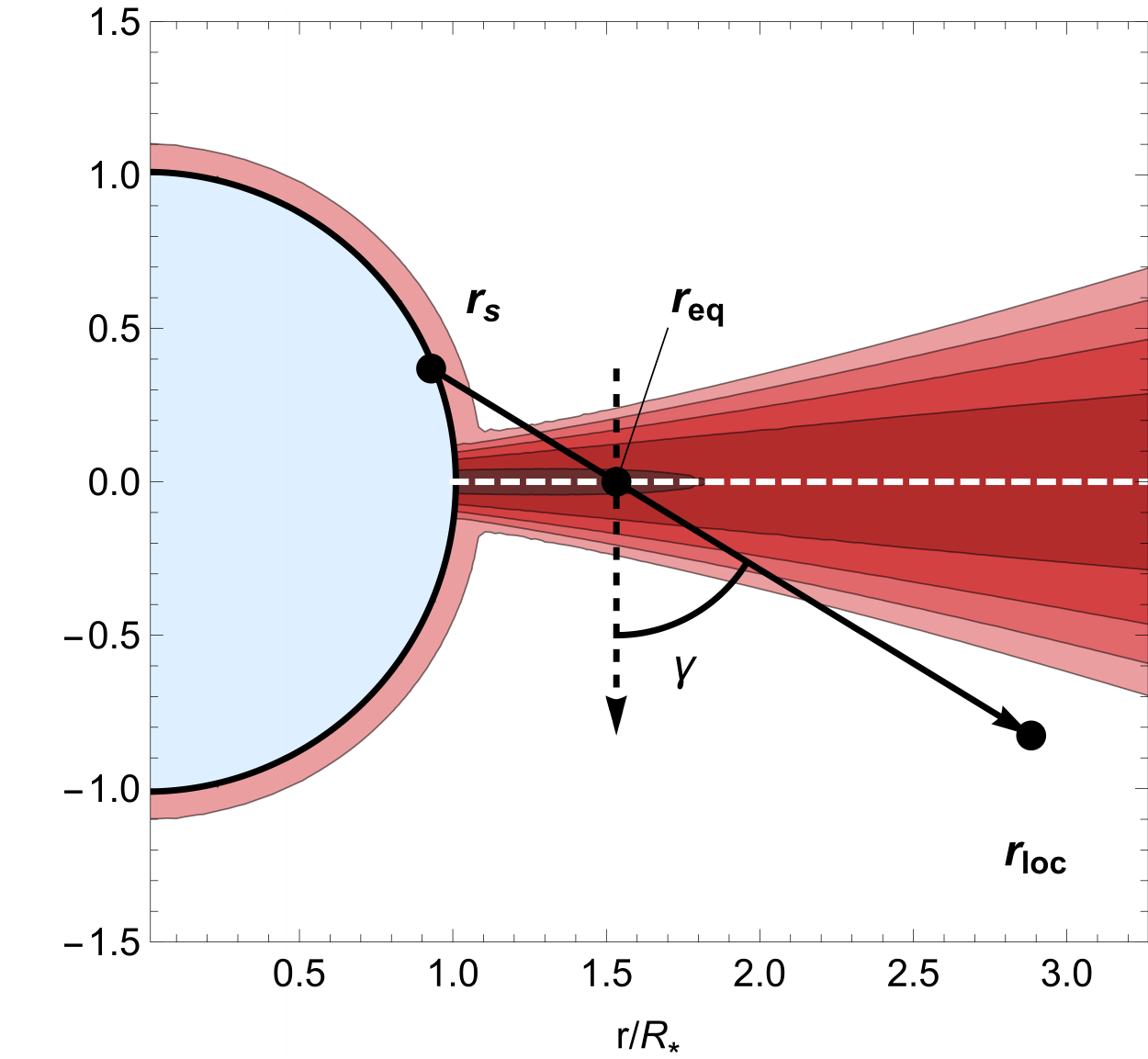}
\caption{\label{fig:thinDiskDiagram} Under the thin disc approximation, optical depth between the stellar surface at $\rs$ and the local point, $\ro$, along the solid black arrow is approximated using the analytic optical depth along the dashed black arrow vertically through the disc at $\req$.}
\end{figure}

For this thin disc approximation, the central assumption is that the radial variations of density and scale height in the accretion disc are much slower than the Gaussian variation of density in the vertical direction.
This means that the dominant contribution to opacity along a ray joining the stellar footpoint at $\rs = \{R_\mathrm{s},z_\mathrm{s},\phi_\mathrm{s}\}$ to a local point in the circumstellar environment at $\ro=\{R_\mathrm{loc},z_\mathrm{loc},\phi_\mathrm{loc}\}$ will arise near $\req=\{R_\mathrm{eq},0,\phi_\mathrm{eq}\}$, where the ray crosses the equatorial plane.
If this is true, then it follows that the primary difference between optical depth along this ray and a ray from $z_\mathrm{s}$ to $z_\mathrm{loc}$ at $R_\mathrm{eq}$ is one of path length.
Explicitly, as diagrammed by figure \ref{fig:thinDiskDiagram}, if we want the optical depth between $\rs$ and $\ro$ (solid line), it can be approximated by the optical depth along the vertical ray (dashed line) corrected for the difference in path length by $1/\cos(\gamma)$, where $\gamma$ is the angle between the two rays.

To test this approximation, we make use of the analytic ablation structure introduced in papers I and II.
Although velocity information is not needed to calculate continuum optical depth, we will make use of the full analytic ablation structure later in this paper, including the velocity structure.
Given additionally the convenience of expressing the wind density as a function of the radial velocity, we therefore chose to begin introducing the analytic ablation structure with this radial velocity component.
As a function of spherical radius $r$ and latitude\footnote{Note that this is in contrast to the \emph{co-}latitude used in standard spherical coordinates.} $\theta$,

\begin{equation}\label{eqn:vr}
v_r(r,\theta)=V_\infty(\theta)\left(1-\delta\frac{R_\ast}{r}\right)\;,
\end{equation}
expressed in terms of the latitude dependent terminal speed, $V_\infty(\theta) \equiv v_r(\infty,\theta)$ and $\delta \equiv 1 - c_s/V_\infty(0)$.
To reproduce the wind regions with terminal speed $V_{\infty,\mathrm{o}}$, the nearly zero radial velocity disc material, and the ablation layer in between,

\begin{equation}\label{eqn:vinf}
V_\infty(\theta)=\frac{V_{\infty,\mathrm{o}}}{2}\mathrm{Erfc}\left(\frac{\theta_\mathrm{d}-\lvert \theta \rvert}{\Delta\theta}\right)\;.
\end{equation}
where $\Delta \theta$ is the half width of the ablation layer, which is centered on latitudes $\pm \theta_\mathrm{d}$.
For consistency with the results we present in section \ref{sec:results}, we chose $\Delta \theta = 4^\circ$ and $\theta_\mathrm{d}=16^\circ$.
To account for disc rotation, we set the azimuthal velocity to be

\begin{equation}\label{eqn:vphi}
v_\phi(r,\theta)=V_\mathrm{orb}(\theta)\sqrt{\frac{R_\ast}{r\, \mathrm{Max}[\cos(\theta),10^{-3}]}}\;,
\end{equation}
where

\begin{equation}\label{eqn:vorb}
V_\mathrm{orb}(\theta) \equiv v_\phi(R_\ast,\theta) = V_\mathrm{orb,o}\left[1-\frac{1}{2}\mathrm{Erfc}\left(\frac{\theta_\mathrm{d}-\lvert \theta \rvert}{\Delta\theta}\right)\right]\;.
\end{equation}

As for density, the analytic ablation structure defines $\rho_\mathrm{eq}$ to be a power-law in cylindrical radius, namely

\begin{equation}\label{eqn:rho_d_eq}
\rho_\mathrm{eq}(R,\phi) = \rho_{\mathrm{d},\mathrm{o}}\left(\frac{R}{R_\ast}\right)^{-\beta}\;,
\end{equation}
with $\beta = 1.5$ chosen to be in the range of pre-main-sequence disc properties cited by \cite{FisHen96}.
The density in the polar regions of the wind\footnote{As mentioned before, in practice this density is so small compared to the disc density that its contribution to the optical depth is effectively zero. Nevertheless, it is included for completeness.} is given by

\begin{equation}
\rho_\mathrm{w}(r) = \frac{\rho_{\mathrm{w},\mathrm{o}}V_\mathrm{\infty,o}}{(r/R_\ast)^2 v_r(r,\theta)}\;,
\end{equation}
and the total density structure is given by the sum

\begin{multline}\label{eqn:rho}
\rho(r,\theta)=\frac{\rho_\mathrm{w}(r)}{2}\mathrm{Erfc}\left(\frac{\theta_\mathrm{d}-\lvert\theta\rvert}{\Delta \theta}\right)\\
+\rho_\mathrm{d}(r\cos(\theta),r\sin(\theta))\left[1-\frac{1}{2}\mathrm{Erfc}\left(\frac{\theta_\mathrm{d}-\lvert\theta\rvert}{\Delta \theta}\right)\right]\;,
\end{multline}
where we have dropped $\phi$ in $\rho_\mathrm{d}$ as the structure is azimuthally symmetric.
For the numerical models of dense disc ablation in this paper, we define the base density of the disc to be

\begin{equation}
\rho_{\mathrm{d},\mathrm{o}} \equiv \frac{1000(\beta-1)}{\kappa_e R_\ast}\;,
\end{equation}
such that the total integrated midplane optical depth, $\tau_\mathrm{d} = 1000$.  
By comparison to the conditions for the simulations in section \ref{sec:results}, we find that $\rho_\mathrm{w,o} \sim 10^{-6} \rho_\mathrm{d}$ and $H(R_\ast) \sim 0.03 R_\ast$.

\begin{figure}
\includegraphics[width=0.5\textwidth]{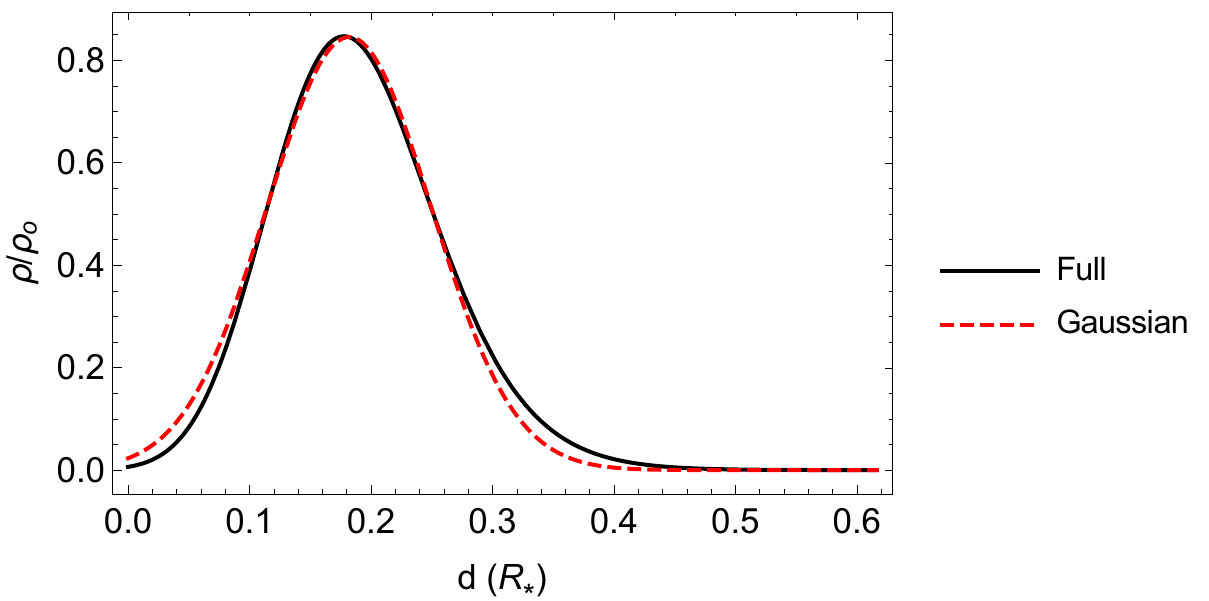}
\caption{\label{fig:rhoComp} For an arbitrarily chosen equator crossing ray proceeding from $r_\mathrm{loc} = 2 R_\ast$, $\theta_\mathrm{loc}=20^\circ$ back toward a point on the star with impact parameter, $b=0.8R_\ast$ and locally centered angle $\phi'=135^\circ$,
a comparison of the actual line of sight density variation of density (black, solid line) against the Gaussian profile (red, dashed line) used in the thin disc approximation, corrected for the difference in path length. Both density profiles are normalized to the disc density at the stellar equator, $\rho_\mathrm{o}$, and lengths are measured in stellar radii.}
\end{figure}

Using this analytic ablation structure, figure \ref{fig:rhoComp} plots the density variation along the ray between an arbitrarily chosen $\rs$ and $\ro$ (solid, black line), and the associated vertical variation (dashed, red line), with the lengths for the vertical density variation stretched by the ratio of path lengths, $1/cos(\gamma)$.
The generally quite good agreement, both in peak density, and in overall spatial variation, between the two density profiles in figure \ref{fig:rhoComp} confirms that the dominant variation along this line of sight arises from the vertical, Gaussian stratification of the disc, and thus that the thin disc approximation is viable.
This in turn allows us to replace $ds$ in equation \ref{eqn:tau_s} with $dz/\cos(\gamma)$.
Plugging in the density profile from equation \ref{eqn:rho_d} allows for the analytic solution,

\begin{multline}\label{eqn:thin_disc_tau}
\tau \approx \sqrt{\frac{\pi}{2}}\kappa_e \rho_\mathrm{eq}(R_\mathrm{eq},\phi_\mathrm{eq}) H(R_\mathrm{eq})\times\\
\left\vert \frac{1}{\cos(\gamma)} \left[\mathrm{Erf}\left(\frac{z_\mathrm{s}}{\sqrt{2} H(R_\mathrm{eq})}\right) - \mathrm{Erf}\left(\frac{z_\mathrm{loc}}{\sqrt{2} H(R_\mathrm{eq})}\right)\right]\right\vert\;,
\end{multline}
where the absolute value is included to ensure the optical depth is always positive.

\begin{figure}
\includegraphics[width=0.5\textwidth]{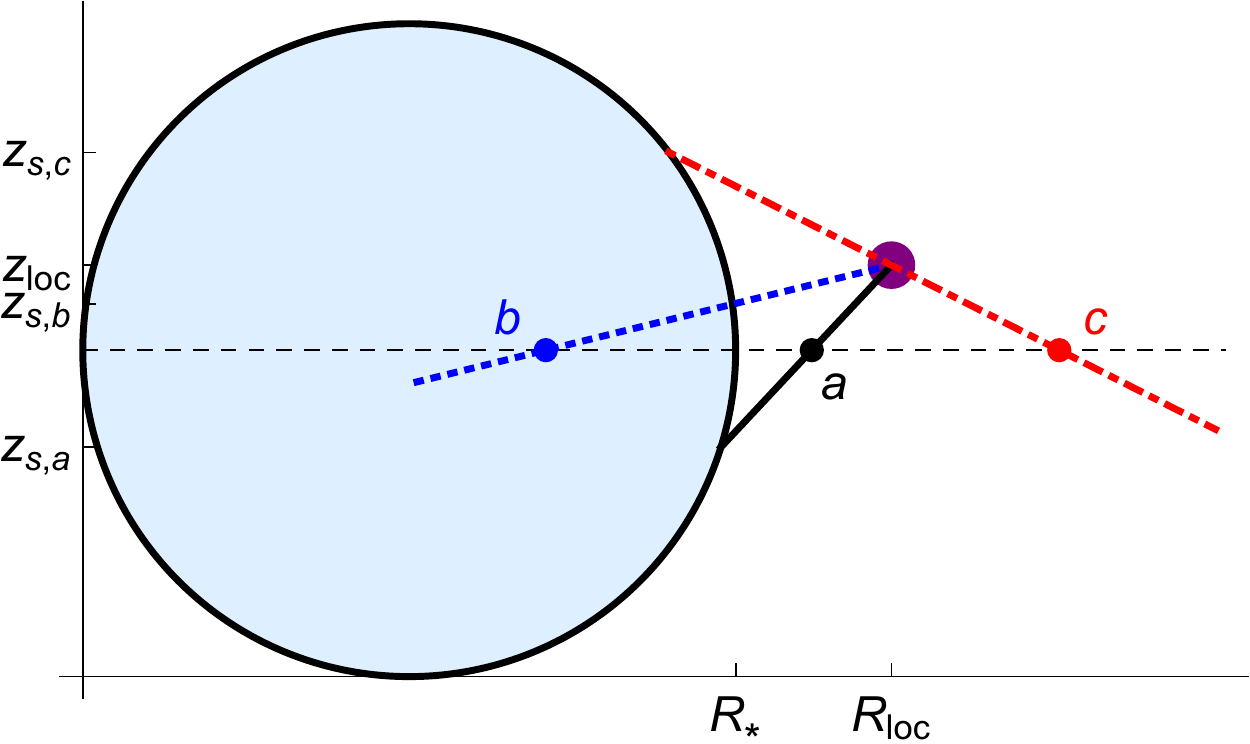}
\caption{\label{fig:cases} Based on the method described here, three cases arise for the analysis. Case a (black, solid line) is the one that the thin disc approximation is designed for, allowing us to take the disc properties at the equatorial crossing point of the ray. In case b (blue, dotted line), we instead take the disc properties at $\{R_\ast,0,\phi_\mathrm{s}\}$. For case c (red, dash-dotted line), we take the disc properties properties directly below the local point, at $\{R_\mathrm{loc},0,\phi_\mathrm{loc}\}$.}
\end{figure}

In order to implement the thin disc approximation in a numerical hydrodynamics code, we then only need to locate the equatorial crossing point of the ray joining $\rs$ and $\ro$, which we can do using vector geometry.
When we do this, depending on the relative positions of $\rs$ and $\ro$, three possible cases arise, as shown in figure \ref{fig:cases}.
The first case, denoted ``a'' and plotted using a solid, black line in figure \ref{fig:cases}, is the one for which the thin disc method is designed.
When $z_\mathrm{s}$ and $z_\mathrm{loc}$ have opposite signs, the equatorial crossing point lies between then on the ray, and we directly take the disc density at this crossing point for $\rho_\mathrm{eq}$ in equation \ref{eqn:thin_disc_tau}.
However, 
when $z_\mathrm{s}$ and $z_\mathrm{loc}$ have the same sign,
one of two things occurs; either $\vert z_\mathrm{s} \vert < \vert z_\mathrm{loc} \vert$ and the equatorial crossing point lies inside, or on the opposite side of, the star (case ``b''; blue, dotted line), or $ \vert z_\mathrm{s} \vert > \vert z_\mathrm{loc} \vert$ and the equatorial crossing point is at a larger radius than $R_\mathrm{loc}$ (case ``c''; dash-dotted, red line).
Given that $\rs$ and $\ro$ can both lie arbitrarily close to the equatorial plane, and thus be highly embedded in the dense disc material, these two cases cannot be ignored.
For case ``b'', when $\vert z_\mathrm{s}\vert < \vert z_\mathrm{loc}\vert$, we take $R_\mathrm{eq} = R_\ast$ and $\phi_\mathrm{eq} = \phi_\mathrm{s}$.
In case ``c'', when $\vert z_\mathrm{s}\vert > \vert z_\mathrm{loc}\vert$, we take $R_\mathrm{eq} = R_\mathrm{loc}$ and $\phi_\mathrm{eq} = \phi_\mathrm{loc}$.
These choices both utilize the disc properties at the closest point on the equatorial plane to either $\rs$ or $\ro$, whichever is closer to the equator, and thus approximate the disc properties for the highest density portion of the ray between $\rs$ and $\ro$.
Note, however, that this case switch only acts to choose the normalization $\rho_\mathrm{eq}$ for equation \ref{eqn:thin_disc_tau}, and that the inclusion of the Erf terms in this equation accounts for the vertical height of the ray away from the disc midplane, which is important for all rays passing only through the disc surface layers.

\subsection{Comparing the Thin Disc Approximation to Long-Characteristic Ray Tracing}
\label{sec:benchmarking}

\begin{figure*}
\centering
\begin{subfigure}{0.32\textwidth}
\includegraphics[width=\textwidth]{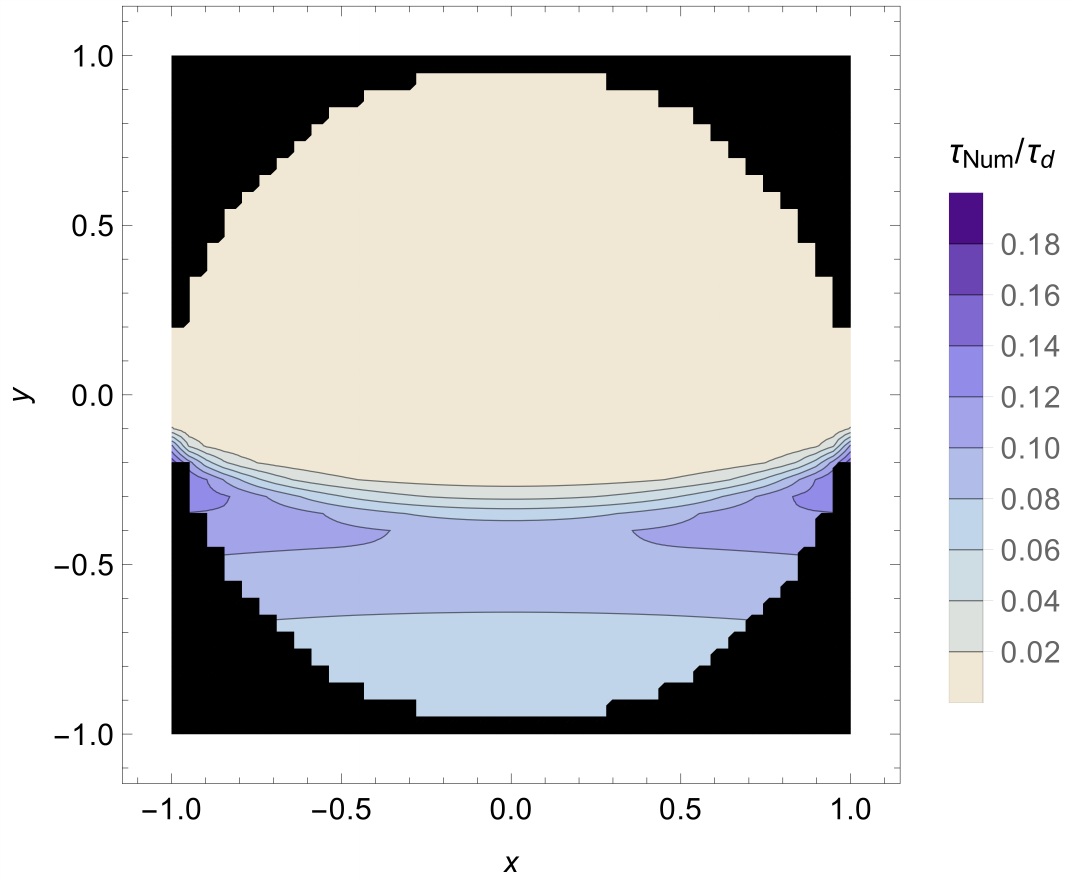}
\caption{Long Characteristics}
\end{subfigure}
\begin{subfigure}{0.32\textwidth}
\includegraphics[width=\textwidth]{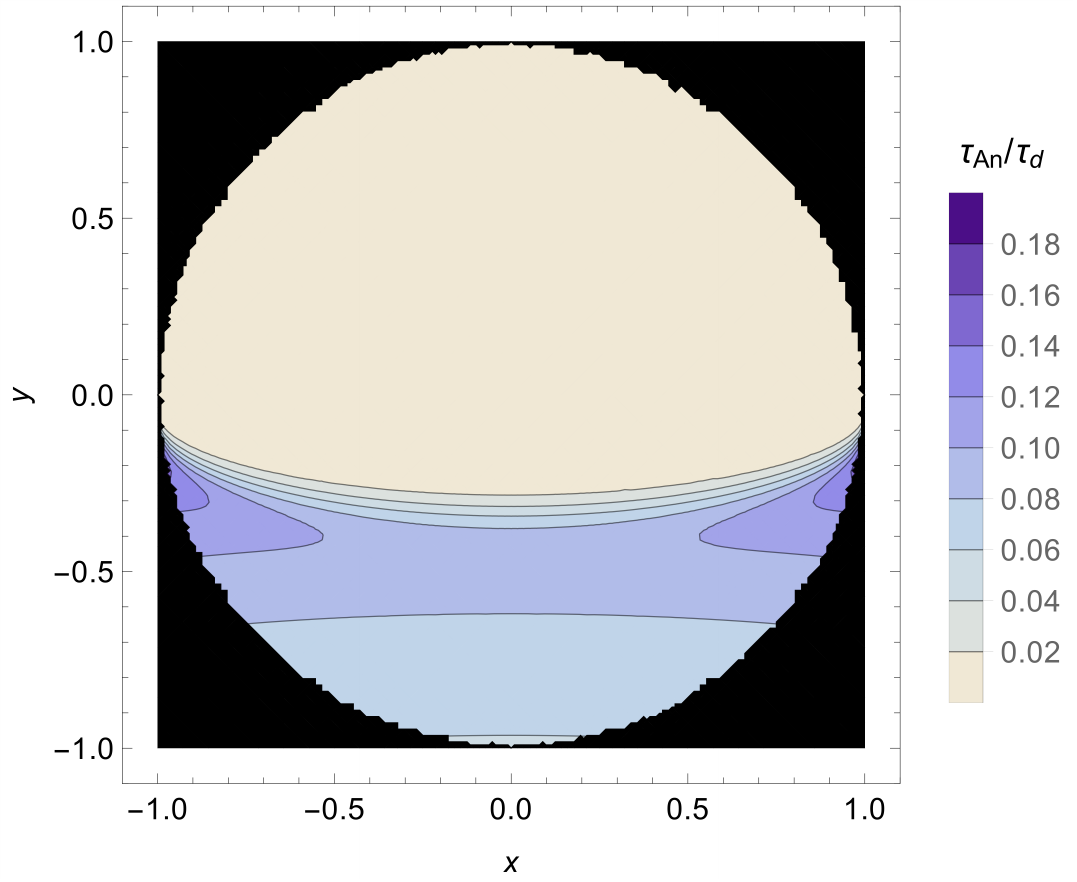}
\caption{Thin Disc Method}
\end{subfigure}
\begin{subfigure}{0.32\textwidth}
\includegraphics[width=\textwidth]{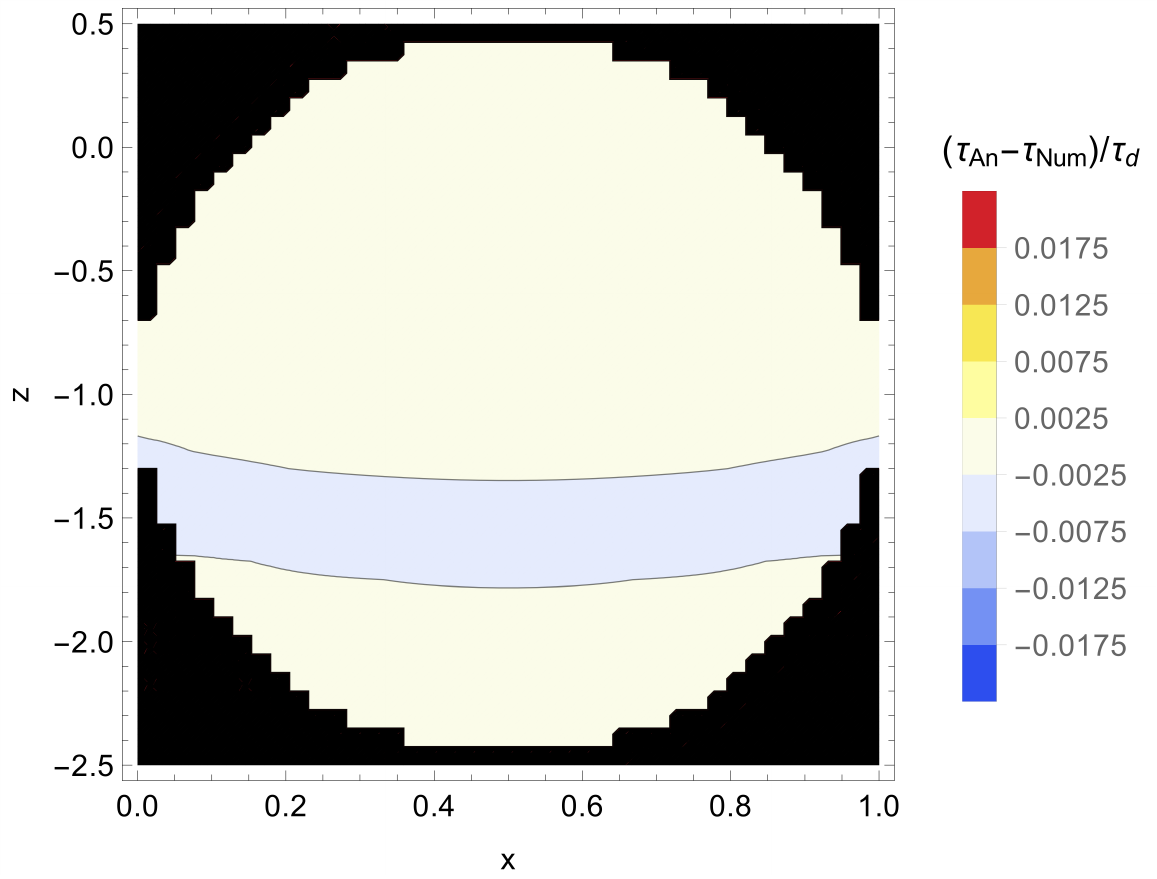}
\caption{Difference}
\end{subfigure}
\caption{\label{fig:tau_onStar_comp} Optical depth as a function of stellar footpoint as seen from a point with local radius $r_\mathrm{loc}=2 R_\ast$ and latitude $\theta_\mathrm{loc}=10^\circ$. This is calculated both using long-characteristic ray tracing (left) and the thin disc approximation (center). The difference between the two methods is plotted in the right panel.}
\end{figure*}

\begin{figure*}
\centering
\begin{subfigure}{0.32\textwidth}
\includegraphics[width=\textwidth]{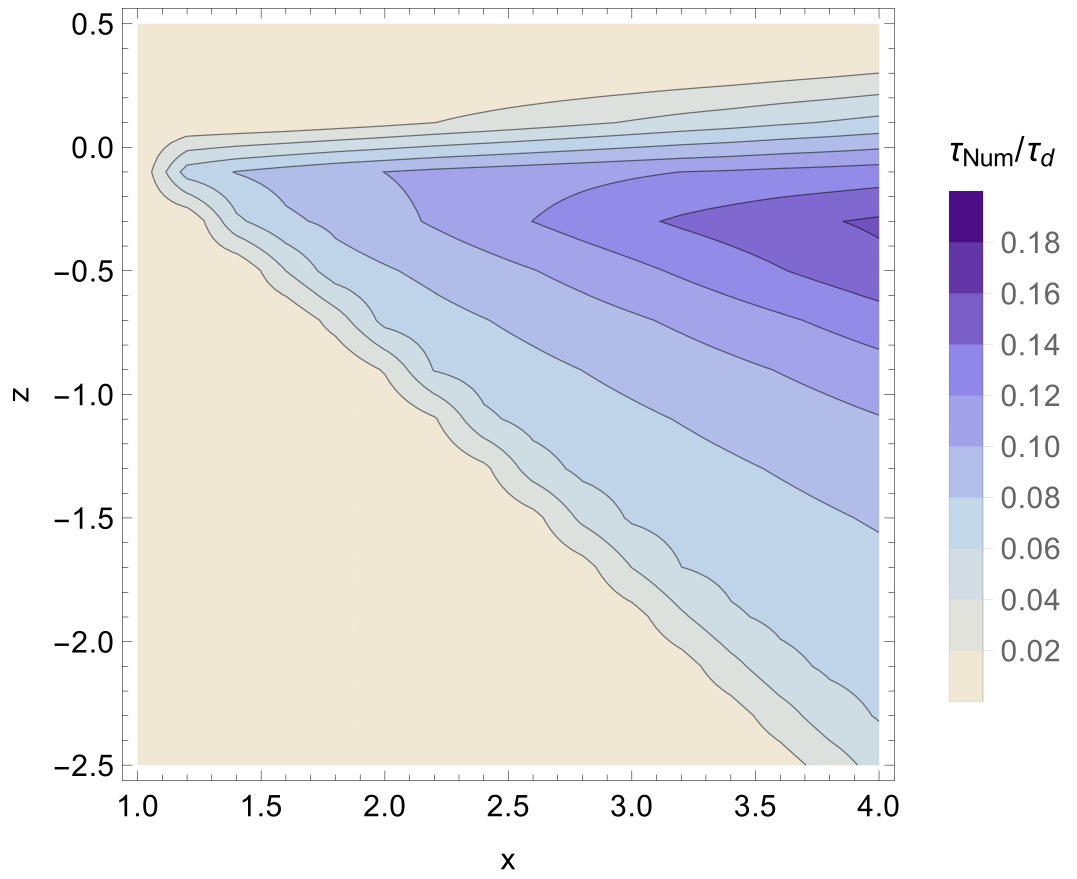}
\caption{Long Characteristics}
\end{subfigure}
\begin{subfigure}{0.32\textwidth}
\includegraphics[width=\textwidth]{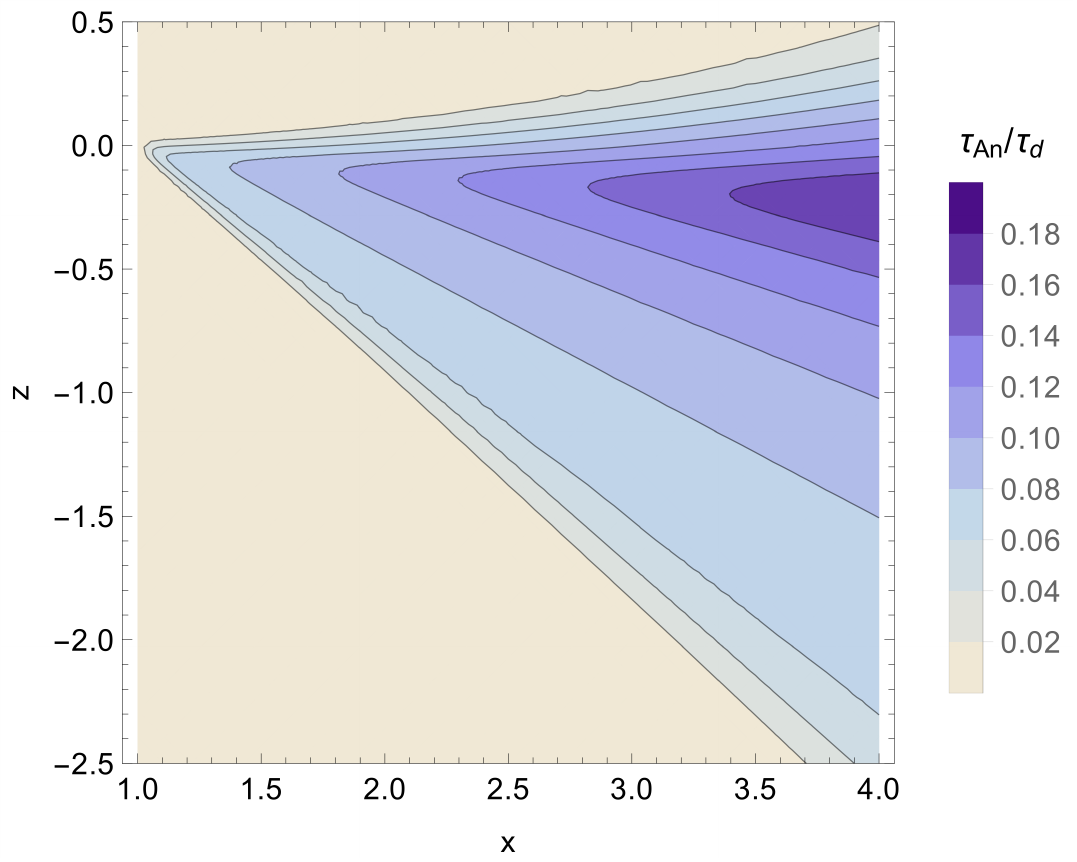}
\caption{Thin Disc Method}
\end{subfigure}
\begin{subfigure}{0.32\textwidth}
\includegraphics[width=\textwidth]{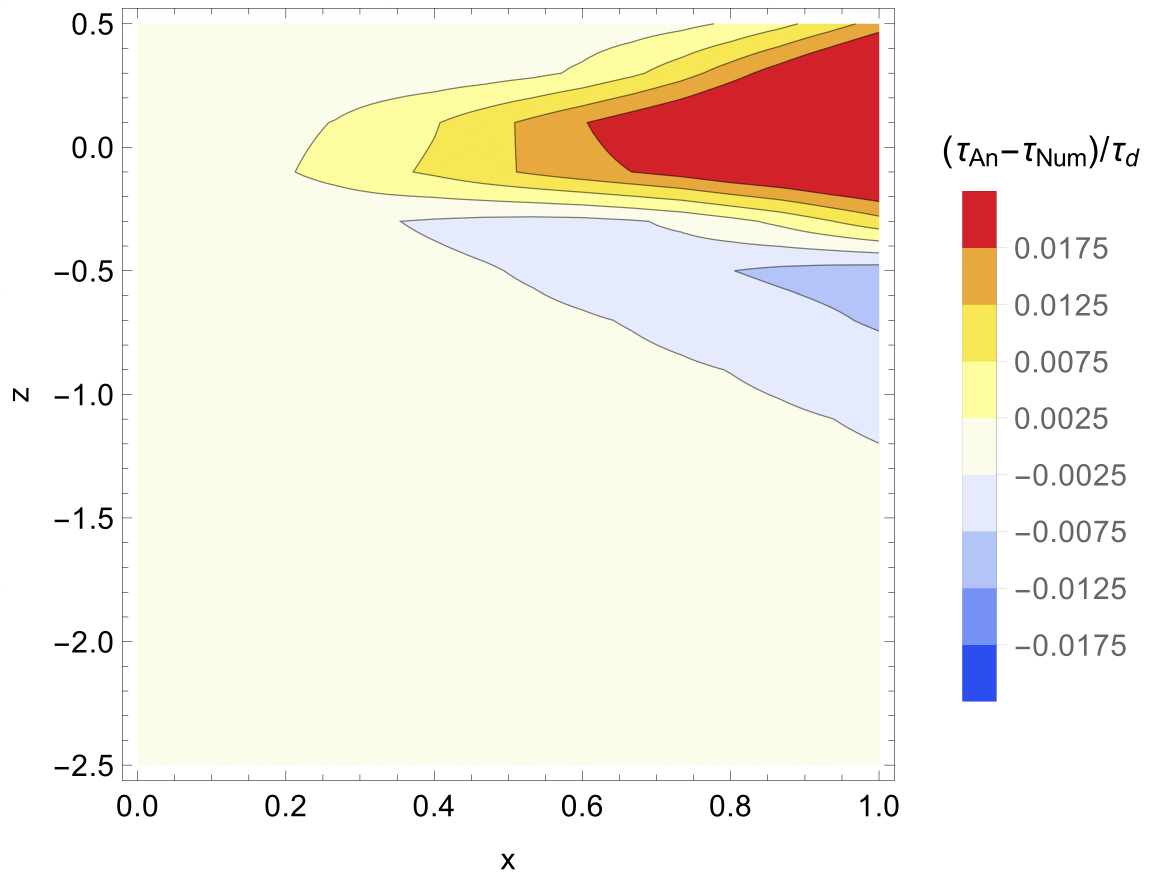}
\caption{Difference}
\end{subfigure}
\caption{\label{fig:tau_XZ_comp} Optical depth as a function of position on the $\{x,z\}$-plane from a point on the star with impact parameter $b=0.8$ and locally-centered angle $\phi'=45^\circ$. Again, this is calculated using long characteristic ray tracing (left) and the thin disc approximation (middle), with the difference plotted on the right.}
\end{figure*}

Before implementing and using the thin disc approximation for hydrodynamics calculations, we test the accuracy with which it reproduces the optical depths.
As a comparison, we use long characteristic ray tracing.
Since both the thin disc approximation and long-characteristic ray tracing can be easily applied to the analytic ablation structure presented in equation \ref{eqn:rho}, this is used for comparison.
Figure \ref{fig:tau_onStar_comp} shows the optical depth between points on the stellar surface and a fixed point at $r_\mathrm{loc} = 2R_\ast$, $\theta_\mathrm{loc}=10^\circ$ using both methods, as well as the difference between the analytic optical depth from the thin disc approximation, $\tau_\mathrm{An}$, and the optical depth numerically determined from ray tracing, $\tau_\mathrm{Num}$.
To supplement this comparison, we also chose a single ray for each local point and calculate the optical depth on this ray.
To chose this ray, we set up a coordinate system at each local point such that $\hat{x}'=\hat{\theta}$, $\hat{y}'=\hat{\phi}$, and $\hat{z}'=\hat{r}$.
Using standard Cartesian to spherical coordinate transformations, we define in this new coordinate system impact parameter $b\equiv r_\mathrm{loc} \sin(\theta')$ and a locally centered angle $\phi'$.
For rays with $b=0.8 R_\ast$ and $\phi'=45^\circ$ the optical depths in the original $\{x,z\}$-plane are shown in figure \ref{fig:tau_XZ_comp}, again with the difference between analytic and numerical optical depths.

Both figures show the optical depth in units of $\tau_\mathrm{d}$, as changing this quantity results in uniformly scaling density, and correspondingly the optical depth, without changing the relative optical depths along different rays.
In general, we find quite good agreement over the majority of the region considered.
Unsurprisingly, the areas with the least good agreement arise when the ray joining $\ro$ to $\rs$ is nearly parallel to the equatorial plane, namely when the ray falls into extreme versions of case ``b'' or ``c'' from figure \ref{fig:cases}. 
For these rays, the underlying assumption of the thin disc approximation that the density variation along a ray arises predominantly from the projection of the vertical, Gaussian variation onto the line-of-sight is no longer correct.
Instead, the projection of the radial gradients of density and optical depth, which are not correctly treated by the thin disc approximation, play an increasing role in the line-of-sight density.
However, such rays are associated with $\ro$ near the equator, where the resulting line acceleration will already be strongly depressed by the high density even in the absence of continuum optical depth effects, as shown in paper I.
Therefore, it would be possible to completely neglect the further reduction from continuum optical depth in this region without impacting the hydrodynamical evolution.
In light of this, and the overall good performance of the thin disc approximation at higher latitudes where UV line-scattering is the dominant force, we proceed with using this approximation in our hydrodynamics simulations.

\section{An Application to Optically Thick Protostellar Discs}
\label{sec:results}

To test the effects of continuum disc opacity on line-driven ablation of optically thick discs, we select two cases, one with a star of $M_\ast=$ 25 $M_\odot$ and one with $M_\ast=$ 50 $M_\odot$.
The associated luminosity of these stars are selected using \cite{EksGeo12}, with mass-radius relation $R_\ast \propto M_\ast^{2/3}$.
The specific stellar parameters used are given in table \ref{tab:stellar_params}, while table \ref{tab:wind_params} presents the parameters for the stellar wind.
First among these wind parameters is the temperature dependent line-acceleration power-law index $\alpha$, initially introduced by \cite{CasAbb75}.
The latter two parameters are the flux- and population-weighted enhancement of line opacity over electron scattering opacity, $\bar{Q}$, and the enhancement above which the assumed distribution of lines is exponentially truncated, $Q_\mathrm{o}$, both introduced by \cite{Gay95}.
The values of all three parameters are obtained from \cite{PulSpr00}.

We initialize the simulations to have discs with $\tau_\mathrm{d} = 1000$, and for each set of initial conditions, we run two simulations, one 
using the thin disc approximation for the line-driven acceleration prescription to account for continuum optical depth effects and one omitting this effect.
All simulations use 256 grid cells in radius from $1-20\;R_\ast$, spaced such that the cell size increases by 2\% with radius between neighboring cells.
In latitude, the simulations use 257 cells evenly spaced in $\theta$ spanning from one pole to the other.
To prevent sound waves propagating through the disc from allowing the outer boundary to feedback on the simulation, we only allow material to flow out of the simulation through the outer boundary in radius.
The inner radial boundary at the stellar surface, however, is set to allow mass flow across it both into and out of the simulation volume, with speeds limited to be at or below the sound speed.
To compute the stellar wind and disc ablation self-consistently, we use the three-dimensional line-acceleration prescription presented in paper I, implemented for these simulations in the (magneto-)hydrodynamics code Pluto \citep{MigBod07,MigZan12}.
We chose to omit the gravitational potential from the disc material, as this is many orders of magnitude lower than that of the star at these length scales.
In order to isolate the effects of continuum optical depth here, we also continue to use isothermal, inviscid simulations, delaying to future work an investigation of the role relaxing these approximations will have.
Note that, by comparison with the results of \cite{HenPul18}, we find that the assumption of isothermality is quite good throughout the layers of the disc where line-driving occurs, such that using the temperature dependent line-driving parameters determined from the stellar effective temperature is a good choice, and we would expect that relaxing the assumption of an isothermal disc would only impact the disc opening angle.
Further, as discussed in more detail later in this section, the ablation layer is launched near the stellar surface where the assumption of isothermality, and thus the disc opening angle, is the best, thus further demonstrating that using an isothermal disc likely has a minor impact on our results.

\begin{table}
\centering
\caption{\label{tab:stellar_params} Stellar Parameters}
\begin{tabular}{c|c|c|c}
\hline
$ M_\ast\;(M_\odot)$ & $R_\ast\;(R_\odot)$ & $L_\ast\;(L_\odot)$ & $T_\mathrm{eff}$ (kK) \\
\hline
25 & 8.5 & $7.34\times 10^4$ & 33\\
50 & 13.6 & $3.52\times 10^5$ & 38\\
\hline
\end{tabular}
\end{table}

\begin{table}
\caption{\label{tab:wind_params} Wind Parameters}
\centering
\begin{tabular}{c|c|c|c|c}
\hline
$ M_\ast\;(M_\odot)$ & $\alpha$ & $\bar{Q}$ & $Q_\mathrm{o}$ &  $\dot{M}_\mathrm{wind}\;(M_\odot/yr)$ \\
\hline
25 & 0.65 & 2200 & 3200  & $5.48\times10^{-8}$\\
50 & 0.66 & 2100 & 2100  & $6.22\times10^{-7}$\\
\hline
\end{tabular}
\end{table}

\begin{figure*}
\centering
\begin{subfigure}{0.45\textwidth}
\includegraphics[width=\textwidth]{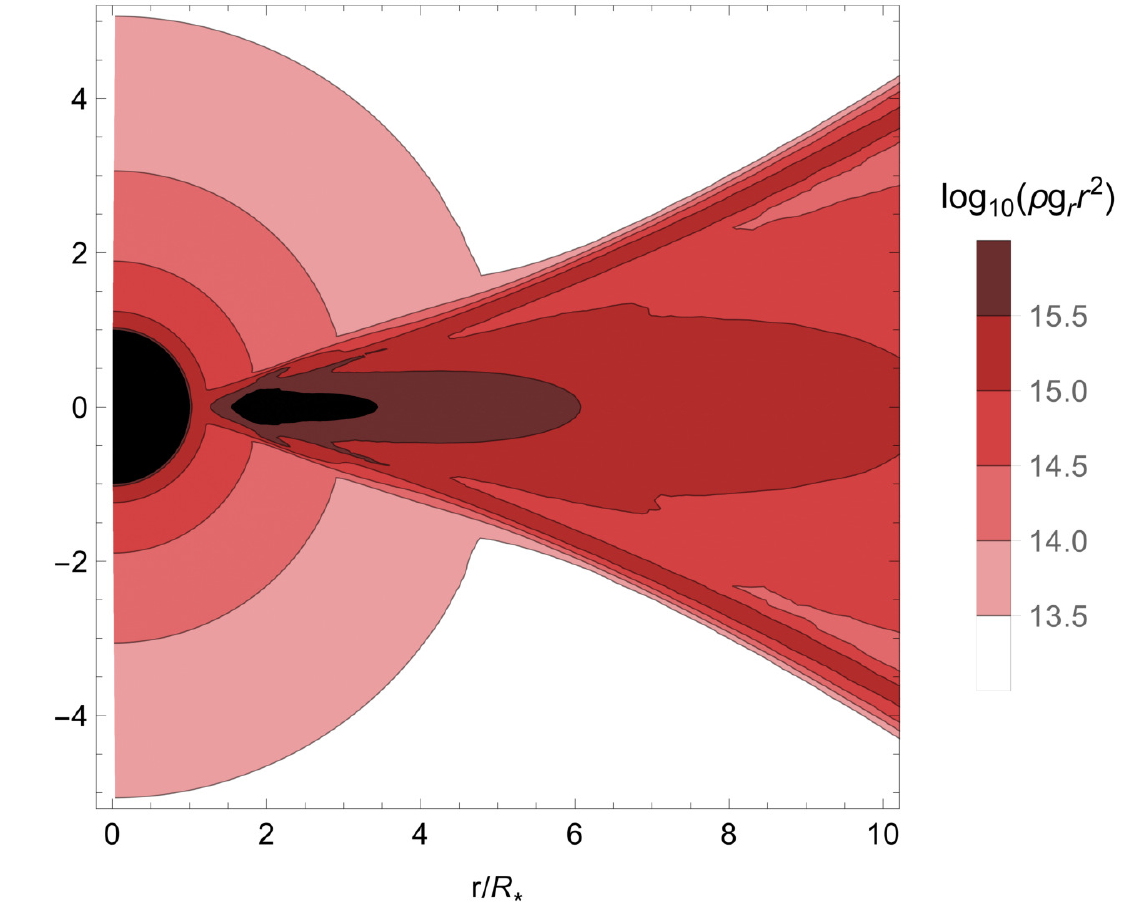}
\caption{No Continuum Opacity}
\end{subfigure}
\begin{subfigure}{0.45\textwidth}
\includegraphics[width=\textwidth]{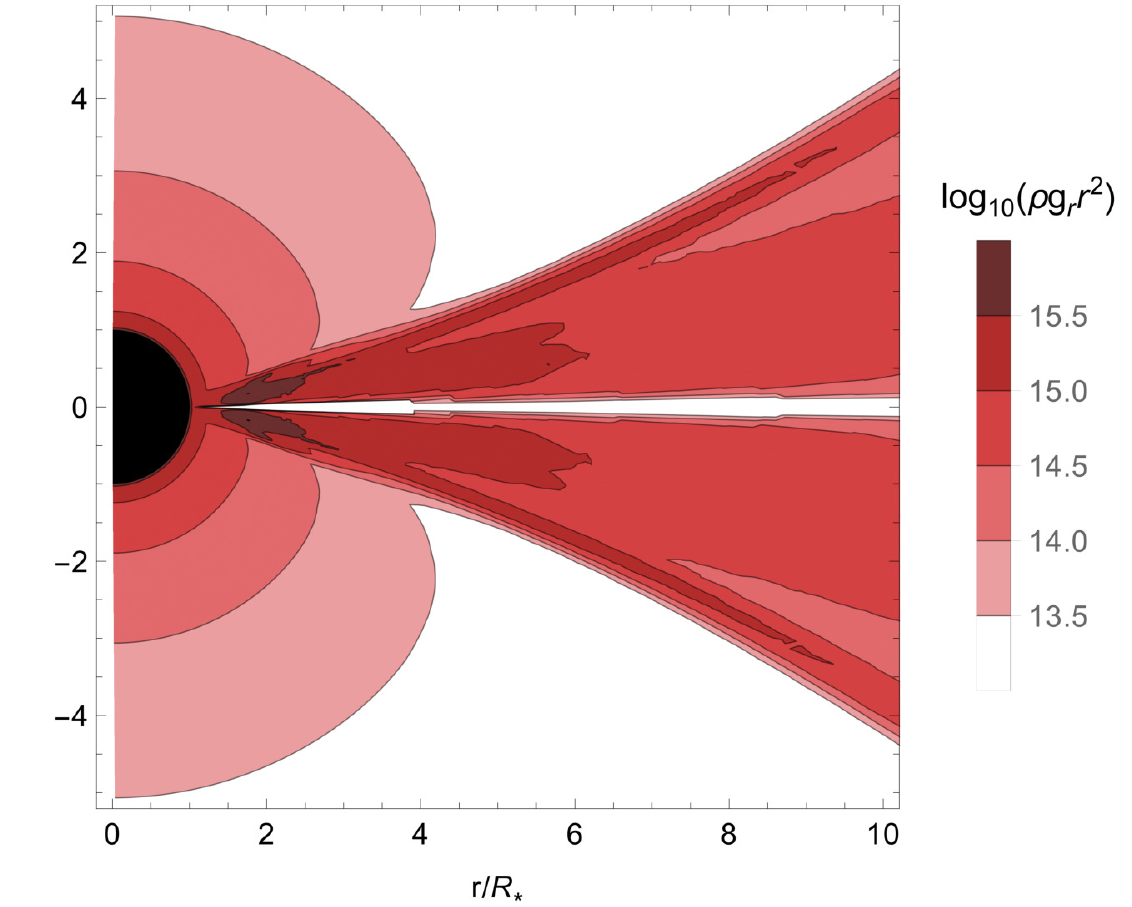}
\caption{Thin Disc Method}
\end{subfigure}
\caption{\label{fig:fpl} Force per unit length averaged over the simulation duration. The left panel omits continuum optical depths while the right panel includes them.}
\end{figure*}

In order to compare the results of the simulations, we first make use of the line-force-per-unit-length, proportional to $\rho g_\mathrm{r} r^2$, where $g_\mathrm{r}$ is the net acceleration in the radial direction.
By examining figure \ref{fig:fpl}, which plots this quantity for both the case omitting continuum opacity (left panel) and the case where we account for the reduction in line acceleration, two interesting morphological features emerge.
First, note that while the force-per-unit-length is completely constant in $\theta$ outside the disc when we omit continuum opacity, including continuum opacity reproduces the expected shadowing by the disc of the wind regions at low latitudes, i.e. $\rho g_\mathrm{r} r^2$ decreases toward the disc surface.
As this shadowing only becomes appreciable away from the stellar surface, it does not decrease the mass loss of the star, as this is set quite close to the stellar surface, but instead reduces the wind terminal speed in these latitudes, resulting in reduced shearing between the wind and disc surface.
However, as noted in paper I, it is the direct action of UV line-scattering forces on the disc surface layers that sets the ablation rate, not shearing of the wind against the disc, so this shadowing should be viewed more as a proof of the efficacy of the thin disc approximation than as an impactful modifier of the behavior of line-driven disc ablation.
 
Second, note that inside the disc, the force-per-unit-length is reduced everywhere by accounting for continuum optical depth.
This reduction is most notable near the equatorial plane where the high optical depth along all rays back towards the star resulting in the line-acceleration actually shutting off completely.
Recalling, however, that line-acceleration is inversely proportional to density, the dynamics of this region of the disc are largely unmodified by 
line-acceleration, even at its full strength in the case omitting optical depth effects.

\begin{figure*}
\centering
\begin{subfigure}{0.45\textwidth}
\includegraphics[width=\textwidth]{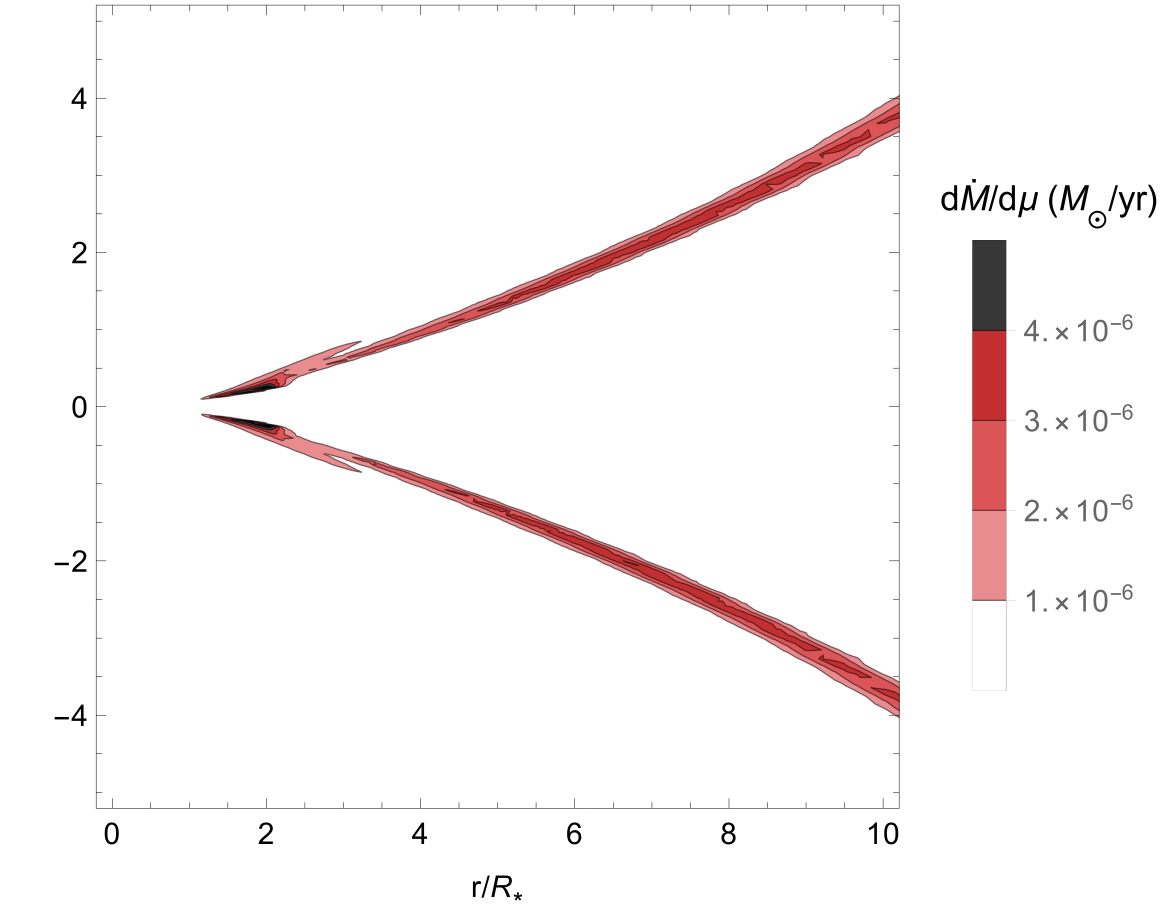}
\caption{No Continuum Opacity}
\end{subfigure}
\begin{subfigure}{0.45\textwidth}
\includegraphics[width=\textwidth]{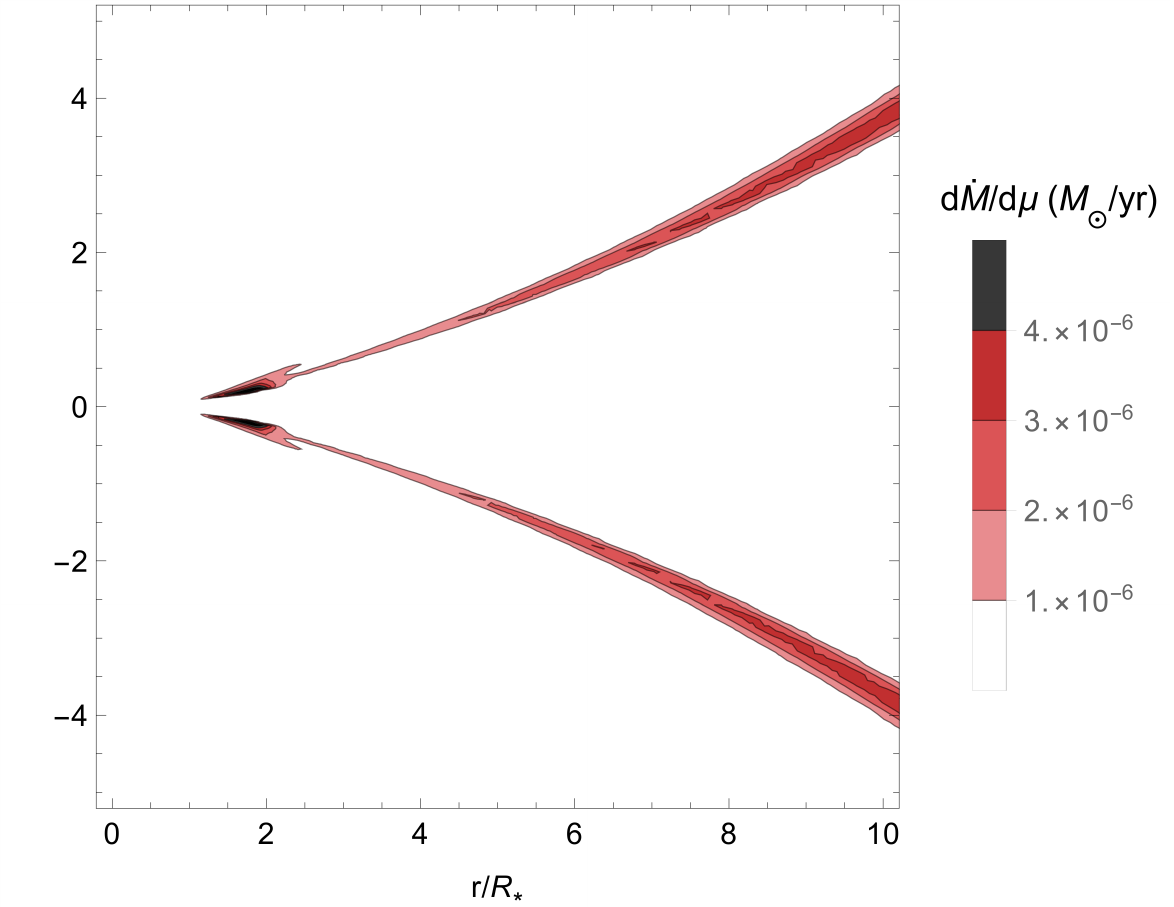}
\caption{Thin Disc Method}
\end{subfigure}
\caption{\label{fig:dMdotdmu} Mass flux per unit solid angle averaged over the simulation duration. The left panel omits continuum optical depths while the right panel includes them.}
\end{figure*}

We can directly see that neither of the notable changes of morphology in force-per-unit-length play an important role in altering the morphology of ablation by considering the mass flux per unit solid angle, for these azimuthally symmetric models proportional to

\begin{equation}
\frac{d\dot{M}}{d\mu}=2 \pi \rho v_r r^2\;,
\end{equation}
where\footnote{Recall that $\theta$ denotes latitude. In standard spherical coordinates with $\theta$ the co-latitude, $\mu = \cos(\theta)$.} $\mu = \sin(\theta)$.
In order to omit the contribution of sound waves excited by the disc/ablation boundary propagating through the disc, we only calculate $d\dot{M}/d\mu$ in regions with supersonic outflows.
Physically, such waves do not contribute to a net loss of disc mass, however, as they average out over time away from the disc surface.
Additionally, we expect that such waves would be damped out in a more realistic viscous disc.
The remarkably close agreement between the left (omitting continuum opacity) and right (including continuum opacity) panels of figure \ref{fig:dMdotdmu}, which plots the time averaged $d\dot{M}/d\mu$ speaks for itself.
Indeed, the only appreciable difference is a slight reduction of the overall level of $d\dot{M}/d\mu$ when continuum opacity is included.

\begin{figure*}
\centering
\begin{subfigure}{0.45\textwidth}
\includegraphics[width=\textwidth]{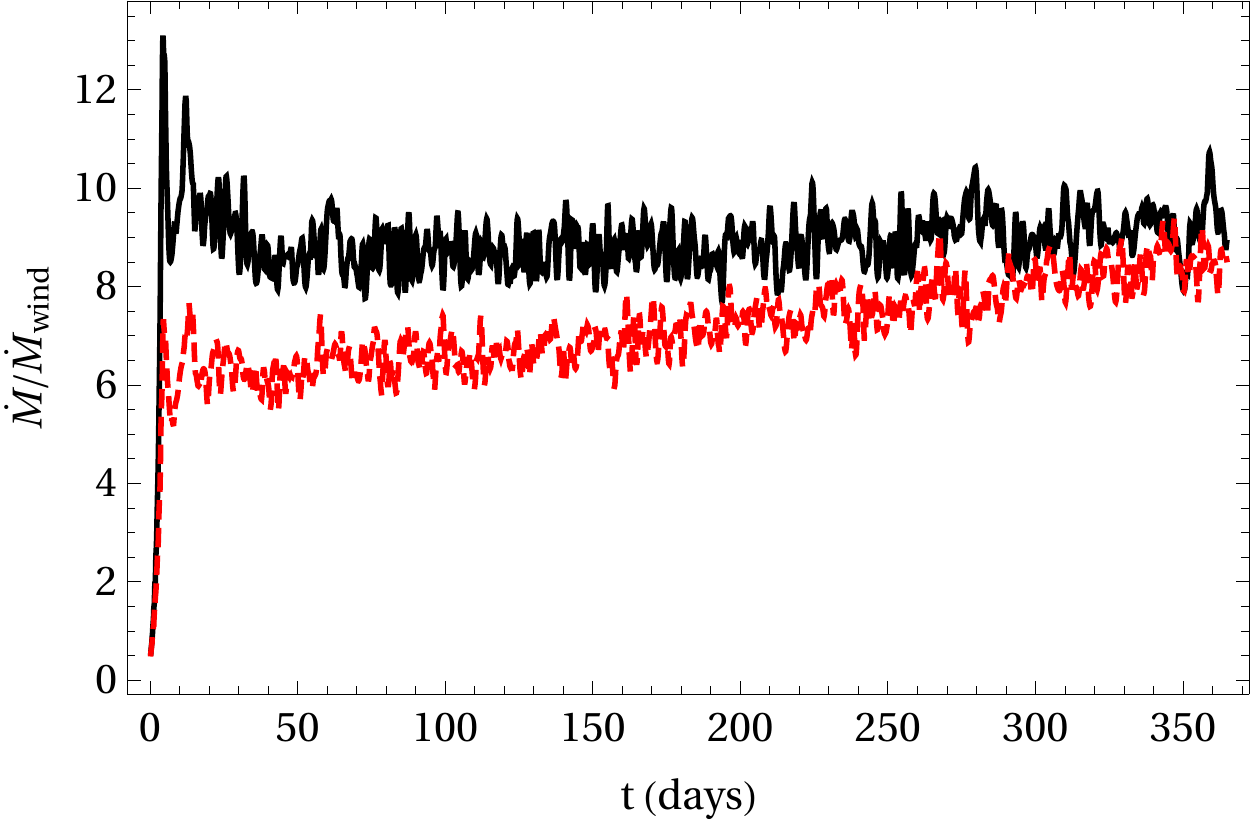}
\caption{25 $M_\odot$ Star}
\end{subfigure}
\begin{subfigure}{0.45\textwidth}
\includegraphics[width=\textwidth]{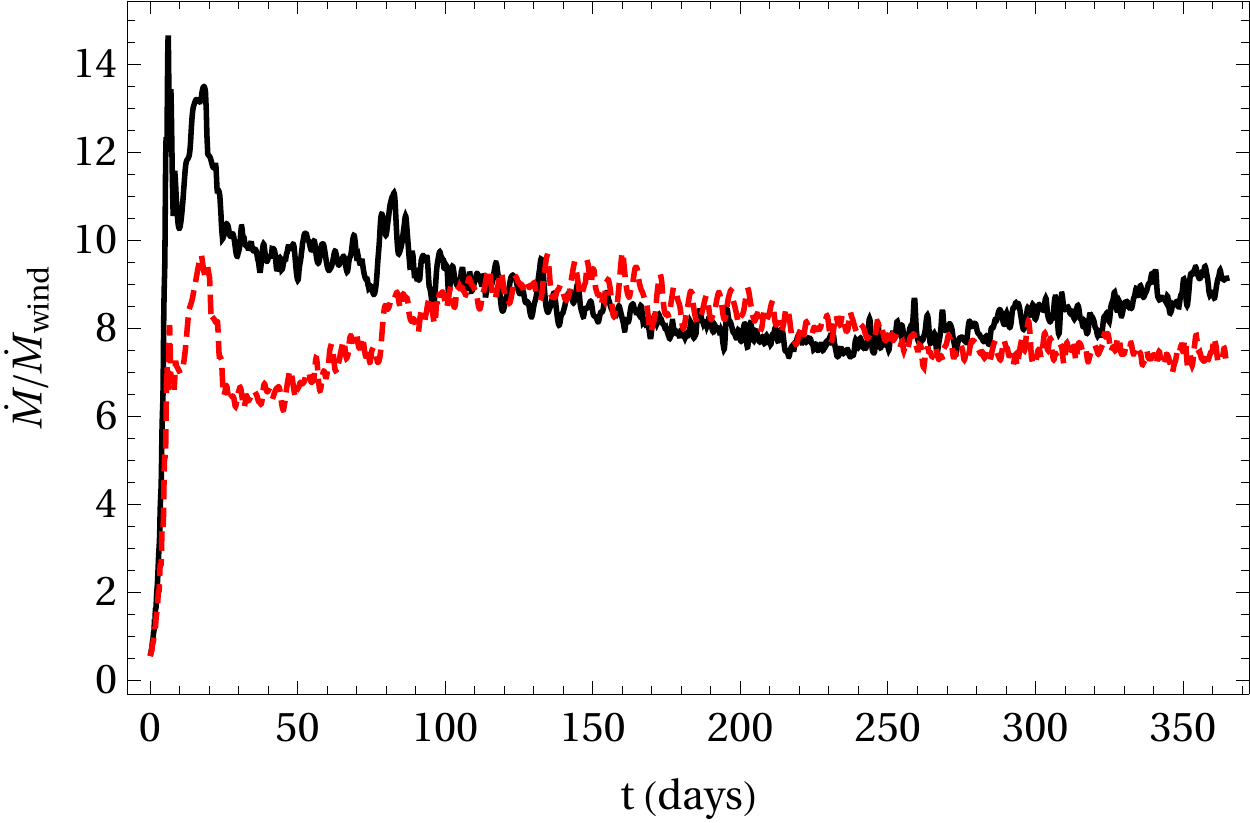}
\caption{50 $M_\odot$ Star}
\end{subfigure}
\caption{\label{fig:CompSuper}  Ablation rate in units of the spherically symmetric wind mass loss rate, $\dot{M}_\mathrm{wind}$, for the simulations with a 25 $M_\odot$ star (left) and 50 $M_\odot$ star (right). The solid black line represents the results from the simulation which ignores continuum opacity, while the dashed red line represents the simulation that includes it.}
\end{figure*}

In order to quantify the level of this reduction in mass loss rate found in figure \ref{fig:dMdotdmu}, we compare the time evolution of the mass loss rate through the outer simulation boundary, i.e. the ablation rate.
We here normalize this ablation rate by the spherically symmetric mass loss rate of the star in the absence of a disc.
These spherically symmetric mass loss rates are provided in table \ref{tab:wind_params}.
For each simulation, we calculate the ablation rate by again only including the contribution of supersonic outflows.
The time evolution of the ablation rate is plotted in figure \ref{fig:CompSuper} for the cases with 25 $M_\odot$ (left panel) and 50 $M_\odot$ stars (right panel). 
In both panels of this figure, the solid black curve denotes the ablation rate from the simulation ignoring continuum opacity, and the dashed red curve the rate calculated from accounting for it using the thin disc approximation.
From these figures, we find that the inclusion of continuum opacity results in a $\sim 30\%$ reduction in ablation rate, and this only in the early phases of the simulations.

\begin{figure}
\includegraphics[width=0.5\textwidth]{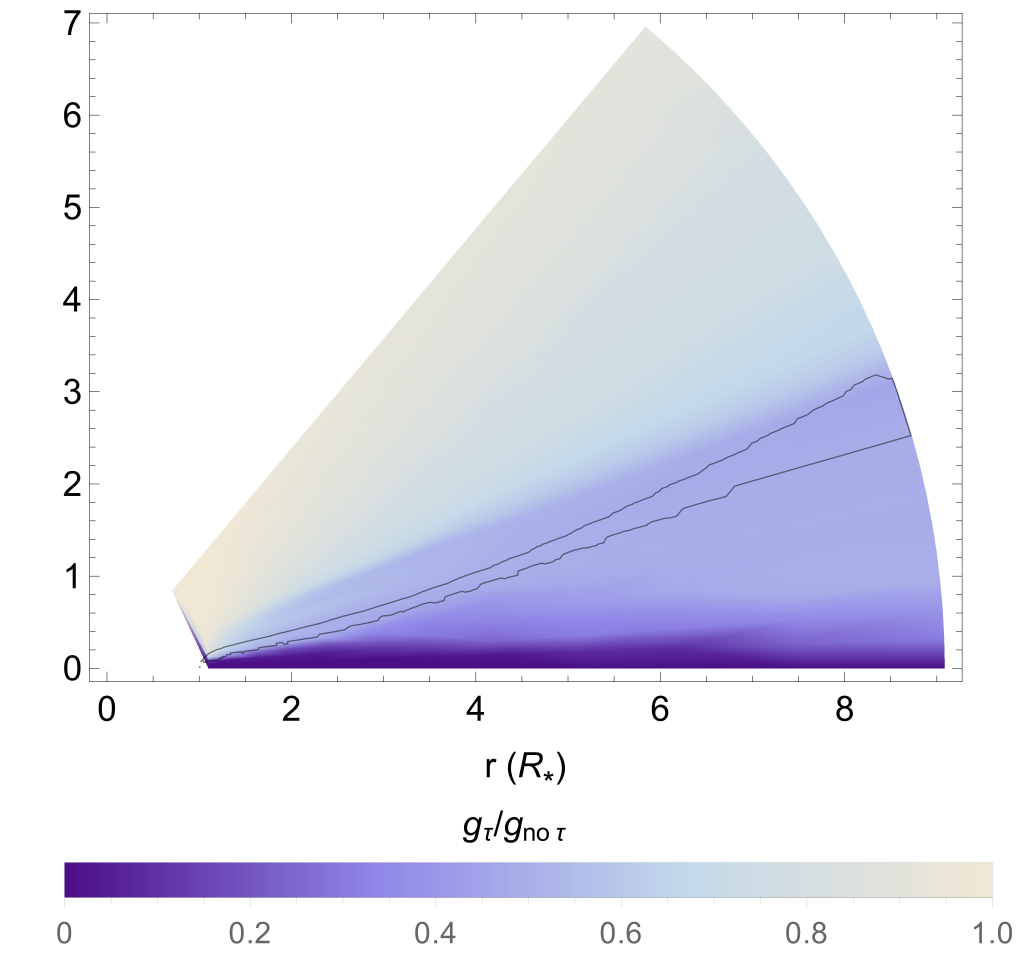}
\caption{\label{fig:gOpacBygNoOpac} For the analytic ablation structure, the ratio of the line acceleration accounting for ($g_\tau$) to omitting ($g_{\mathrm{no}\;\tau}$) continuum optical depth. The single contour outlines the ablation layer from the simulation using a 25 $M_\odot$ star and omitting continuum opacity.}
\end{figure}

To understand the origin of this relatively small reduction, we calculate line acceleration as a function of position from the analytic ablation structure.
Figure \ref{fig:gOpacBygNoOpac} plots the ratio of line acceleration for the case where we account for continuum opacity to the case where it is omitted.
The single contour in this figure, included to give a general idea of the latitudes where ablation occurs, shows the location of the ablation layer for the snapshot at $t=5$ days of the simulation using a 25 $M_\odot$ star and omitting continuum opacity.
As expected, the closer to the equatorial plane a point is, the larger the fraction of the star is obscured, and the more strongly line acceleration is reduced.
With the exception of latitudes deeply inside the disc, it is also the case that the farther from the star a point is, the more optical depth reduces line acceleration, again because more of the star becomes obscured by the disc.
For points in the ablation layer, for instance, the reduction goes from negligibly small to order $50\%$ with increasing distance from the star.
The detailed origin of the $\sim 30\%$ reduction in ablation rate, is thus likely a tracer of the conditions at the location where ablation is launched, which, based on figure \ref{fig:gOpacBygNoOpac}, reinforces the suggestion from paper I that ablation is launched quite close to the inner rim of the disc.

As we are already discussing figure \ref{fig:gOpacBygNoOpac}, it is worth investigating the behavior that deep inside the disc there are regions where moving away from the star results in less reduction of the line acceleration before continuing the discussion on ablation rate.
Careful investigation reveals that this behavior is the result of the decreased accuracy of the thin disc approximation when rays joining $\rs$ to $\ro$ become close to parallel to the equatorial plane, as noted in section \ref{sec:benchmarking}.
At large radii, these rays can end up in extreme versions of cases ``b'' and ``c'' from figure \ref{fig:cases} as the solid angle subtended by the star shrinks, and any rays impacting it are already nearly radial.
Moreover, switching from case ``b'' to ``c'' has an increasingly large effect as $r_\mathrm{loc}$ increases, due to both $\rho_\mathrm{eq}$ and $H$ being calculated at the stellar surface in case ``b'' and at $R_\mathrm{loc}$ in case ``c''.
This behavior reinforces that the thin disc approximation is designed to handle cases where the primary variation of density along the ray is due to the vertically Gaussian stratification of the disc rather than the radial dependence of $\rho_\mathrm{eq}$ or $H$, and more specifically cases where the ray considered passes through the equatorial plane relatively close to $R_\mathrm{loc}$.
As mentioned before, however, the region where this condition breaks down is embedded in the disc away from the stellar surface and inner edge of the disc.
Here line-acceleration is already several orders of magnitude sub-dominant due to the high density, and comparison to the simulations omitting continuum opacity show that even ignoring optical depth effects in this region does not change the dynamics.

Returning now to the ablation rate, we examine the general trend of ablation rate in the presence of continuum opacity to approach ablation rate ignoring continuum opacity as the simulation proceeds.
This, however, is unsurprising as it is simply a result of the disc not being replenished.
Due to our choice to omit viscous accretion for these simulations, the ablation is always able to detach the disc from the star eventually.
As the disc is detached from the star, it obscures a decreasing fraction of the star as seen from the regions where ablation is launched, and thus continuum optical depth of the disc becomes of decreasing importance.
Specifically, for the simulation with a 25 $M_\odot$ star including continuum opacity, the final snapshot of the simulation shows the disc to be evacuated to $\sim 0.8 R_\ast$ away from the stellar surface.
For the simulations with a 50 $M_\odot$ star, the wind mass loss rate and ablation rate are approximately an order of magnitude higher than in the simulations with a 25 $M_\odot$ star, causing the disc to be evacuated out to the same relative distance of $0.8 R_\ast$ after only about 100 days, and thus explaining the convergence of ablation rate in the simulations with and without continuum optical depth on this timescale.

Finally, before moving on to the summary of this work, it is worth re-emphasizing that this $\sim 30\%$ is the maximum expected reduction from continuum optical depth effects.
The disc is high enough density that the continuum electron scattering optical depth on rays through the disc is on the order of several tens, so for practical purposes intensity on these rays is zero, and thus this study already treats the situation with maximal reduction in stellar intensity from continuum optical depth effects.
Additionally, recall that all calculations here are done under the assumption of a pure absorption opacity, i.e. we have omitted the scattering source function.
Taking continuum electron scattering to be the dominant source of opacity in the UV for the conditions around a forming massive star, correctly implementing a scattering source function would result in ablation rates between those found in the two limiting cases we treat here, and thus a yet more modest reduction in ablation rate.

\section{Summary and Future Work}
\label{sec:summary}

We have here presented the thin disc approximation as a method for treating continuum optical depths in simulations of star-disc systems.
We find that this method generally performs quite well in estimating optical depths through high density discs, especially near the stellar surface.
Given that the regions where we find that it performs less well are of limited interest for the problem we are interested in, namely line-driven ablation of high mass discs around forming massive stars, we implement and utilize this method in numerical hydrodynamics simulations.
Subsequent analysis of these simulations, however, demonstrates that accounting for continuum optical depth results in a $\sim 30\%$ reduction in ablation rate from highly optically thick discs.
Recalling that the optical depths implemented assume a pure absorption opacity, the reduction in ablation rate from continuum optical depth effects is likely even more modest than this $30\%$.

Future natural outgrowths of this work fall into two major categories; 1) improvements in accounting for optical depth effects and 2) studies of ablation from star forming discs.
In terms of future potential treatments of optical depth effects, a directly possible next step would be addressing the role of scattering radiative transfer in line-driven ablation.
Indeed, the first steps toward such a study are already taken by \cite{HenPul18}, which self-consistently computes the radiation temperature profile for the analytic ablation structure introduced in paper II and reiterated in section \ref{sec:thin_disc}. 
Extending such a study to additionally calculate the line-acceleration could naturally address both the effects of multiple line resonances and of continuum disc opacity in a single study.
However, given the complex nature of treating multiple resonances, and of obtaining a self-consistent scattering source function for the radiative transfer, it is likely that such a study would be limited to snapshots of the density and velocity structures.
Given the conclusions of paper II on the important role of off-star resonances in filling back in intensity and line-acceleration lost from on-star resonances, the contribution of scattering from the disc surface could prove to have interesting implications for line-driven disc ablation.

Another potential improvement over the thin disc approximation could be a method that treats the non-isothermality of accretion discs, and the resulting set of opacity contributors.
While this was not necessary for the study presented here, due to our interest in optical depths in the UV, such a model could be of interest for studies which additionally or instead treat the role of radiation pressure on molecules or potentially even dust grains in the very deep layers of the disc away from the star.

In terms of studies of line-driven ablation of accretion discs in massive star formation, there is a key point of interest embedded in the results of this paper that has thus far gone un-addressed.
Ablation in these simulations occur at a ten times higher rate from these high density discs than from the lower density classical Be star discs we have studied thus far.
While this paper has been focused on the potential role of continuum optical depth effects, future work should prioritize a more detailed study of the origins of this enhanced ablation rate and its dependence on stellar and disc parameters.
Such a study could prove to be highly interesting in better constraining the nature of the final miles of accretion from au length scales down to the stellar surface.

\section*{Acknowledgements}

This study was conducted within the Emmy Noether research group on ``Accretion Flows and Feedback in Realistic Models of Massive Star Formation" funded by the German Research Foundation under grant no. KU 2849/3-1.
Simulations were carried out using the computational resources available through the bwHPC initiative for high-performance computing in
Baden-W\"urttemberg, Germany.
We thank the anonymous referee for providing helpful comments that enhanced and clarified the discussion in this paper.

\bibliographystyle{mn2e}
\bibliography{biblio}

\end{document}